\documentclass[onecolumn,letterpaper,11pt,draftclsnofoot]{IEEEtran}

\usepackage{algorithm,algorithmic,amsbsy,amsmath,amssymb,epsfig}

\DeclareMathAlphabet{\mathpzc}{OT1}{pzc}{m}{it}
\usepackage{times,amsmath,epsfig}
\usepackage{cite}
\usepackage{graphicx}
\usepackage{psfrag}
\usepackage{subfigure}
\usepackage{url}
\usepackage{stfloats}
\usepackage{amsmath}
\usepackage{amssymb}
\usepackage{array}
\usepackage{array}
\usepackage{slashbox}

\newcommand{\beq}{\begin{equation}}
\newcommand{\eeq}{\end{equation}}
\newcommand{\bitm}{\begin{itemize}}
\newcommand{\ba}{\begin{array}}
\newcommand{\ea}{\end{array}}
\newcommand{\eitm}{\end{itemize}}
\newcommand{\beqn}{\begin{eqnarray}}
\newcommand{\eeqn}{\end{eqnarray}}
\newcommand{\beqno}{\begin{eqnarray*}}
\newcommand{\eeqno}{\end{eqnarray*}}
\newcommand{\bma}{\begin{displaymath}}
\newcommand{\ema}{\end{displaymath}}
\newcommand{\bnu}{\begin{enumerate}}
\newcommand{\enu}{\end{enumerate}}
\newcommand{\bce}{\begin{center}}
\newcommand{\ece}{\end{center}}
\newcommand{\btb}{\begin{tabular}}
\newcommand{\etb}{\end{tabular}}

\begin{document}
%

\title{Resource Allocation in Full-Duplex Communications for Future Wireless Networks}
\author{
\IEEEauthorblockN{Lingyang Song\IEEEauthorrefmark{1}, Yonghui Li\IEEEauthorrefmark{2}, and Zhu Han\IEEEauthorrefmark{3}\\}
\IEEEauthorblockA{\IEEEauthorrefmark{1}School of Electrical Engineering and Computer Science, Peking University, Beijing, China}
\IEEEauthorblockA{\IEEEauthorrefmark{2}School of Electrical and Information Engineering, The University of Sydney, Australia}
\IEEEauthorblockA{\IEEEauthorrefmark{2}Electrical and Computer Engineering Department, University of Houston, USA}
}

\maketitle

\thispagestyle{empty}
\begin{abstract}
The recent significant progress in realizing full-duplex~(FD)
systems has opened up a promising avenue for improving quality of service (QoS) and quality of experience (QoE) in future wireless networks. There is an urgent need to address the diverse set of challenges regarding different aspects of
FD network design, theory, and development. In addition to the
self-interference cancelation signal processing algorithms, network
protocols such as resource management are also essential in the
practical design and implementation of FD wireless networks. This
article aims to present the latest development and future directions
of resource allocation in different full duplex systems by exploring
the network resources in different domains, including power, space,
frequency, and device dimensions. Four representative application
scenarios are considered: {FD MIMO networks},  {FD
cooperative networks}, {FD OFDMA cellular networks}, and
{FD heterogeneous networks}. Resource management problems and
novel algorithms in these systems are presented, and key open
research directions are discussed.

\end{abstract}


\section{Introduction}

With more and more new multimedia rich services being introduced and
offered to a rapidly growing population of global subscribers, there
is an ever-increasing demand for higher data rate wireless access,
making more efficient use of the precious resource a
crucial need. As a consequence, new wireless technologies such as
Long Term Evolution~(LTE) and LTE-Advanced have been introduced.
These technologies are capable of providing high speed, large
capacity, and guaranteed quality-of-experience~(QoE) mobile
services~\cite{Song2010,Dong2014}. However, all existing wireless
communication systems deploy the half-duplex~(HD) radios which
transmit and receive the signals in two separate/orthogonal
channels. They dissipate the precious resources by employing either
a time-division or a frequency-division duplexing. Though a
full-duplex~(FD) system, where a node can send and receive the
signals at the same time and frequency resources, offers the
potential to double the spectral efficiency, for many years it has
been considered impractical. This is because the signal leakage from
the local output to input  may overwhelm the receiver, thus making it impossible to extract the
desired signal.

Recently, there has been a significant process in the self
interference cancelation in FD systems. Depending on the size of
devices, it has been shown that the amount of self-interference and
the extent to which it can be canceled varies greatly, thus
impacting the performance of different nodes in the network
differently. In \cite{Melissa}, several interference cancelation
mechanisms have been proposed. It was shown experimentally that it
is possible to adequately reduce the self interference to a certain
level at which the FD radios achieve a higher rate than the HD
systems. In \cite{fmradios}, new analog and digital cancelation techniques were
developed and implemented in an in-band FD WiFi radios. It is shown
experimentally that the self interference level can be reduced to
the receiver noise floor. These significant progresses in hardware
design and signal processing techniques have presented a great
potential for realizing the FD communications in a near future for
the next generation cellular networks.

Since the FD technology enables to explore another dimension of the
network resources to increase the network capacity, it requires the
new design of network protocols and resource allocation algorithms
in FD communications systems. This promising opportunity has so far
inspired the rapid research development in this area.
In~\cite{XiZhang}, the optimal power allocation among the FD source
nodes was presented to maximize the sum-rate of wireless FD
bidirectional transmissions. In~\cite{Emadi}, the optimum power allocation
schemes subject to individual power constraints have been
analytically obtained for a FD decode-and-forward relay channel.
In~\cite{Riihonen2011TWC}, the gain factor of amplify-and-forward
relaying was optimized to maximize the signal-to-interference and
noise ratio~(SINR) and at the same time prevent the oscillation
effects at the relay caused by the residual interference.

Though resource management is essential to system performance, most
existing work mainly focuses on power allocation in FD wireless
networks. Actually, many other network resources in space,
frequency, and device dimensions can be explored in FD networks to
further reduce the self interference and at the same time improve
the system spectrum efficiency. In this regard, there is a
significant need to address the various challenges in the theory,
design, and development of FD systems. In this article, we
comprehensively discuss the novel resource allocation algorithms for
FD communication systems to optimize their network performance. In
particular, we focus on the following major application scenarios:
\begin{itemize}
  \item FD MIMO systems (FD-MIMO): Each node in the FD-MIMO systems is equipped with a FD radio and multiple antennas, each of which can be used for transmission and reception. This enables a simultaneous bidirectional information exchange between two nodes. The resource allocation in such systems involves the allocation of spatial domain resources, such as antennas~\cite{zhou2014, zhou2015}.
  \item FD relay networks (FD-Relay): The basic FD-Relay network structure consists of one source and destination pair, and one relay node. Both the source and destination nodes are HD, but the relay is operated in the FD mode. The resource allocation in such systems involves the allocations of antennas, relays and power~\cite{Baranwal2015,cui2014,Suraweera2014,yang2014,aazhang2013}.
  \item FD OFDMA networks (FD-OFDMA): It is composed of a FD base station~(BS) using OFDM, and multiple single antenna HD uplink and downlink users. The uplink and downlink users can form a transmit-receive pair to communicate with the FD BS. The key challenge in such a network is how to optimally pair the uplink and downlink users for each OFDM subcarrier in communicating with the FD BS~\cite{di2014}.
  \item FD heterogenous networks (FD-HetNet): In such a network, the macro base stations (MBSs) and femto access points (FAPs) are equipped with FD radios. In contrast to the traditional HetNet systems where FAPs in the adjacent cells typically do not interfere much with each other, in FD-HetNet the simultaneous uplink and downlink communications between the users and MBSs/FAPs can lead strong interference among the FAPs~\cite{Sultan2014}.
\end{itemize}

Obviously, in different FD application scenarios, different network
resources need to be optimized by exploring different resource
allocation algorithms. In this magazine paper, we demonstrate, in
the above mentioned FD application scenarios, new research
challenges in resource allocation and network protocols, and present
the latest promising research development to resolve these technical
challenges. Some potential research directions and open problems
will be also discussed.

The rest of article is organized as follows: Section~II reviews the
basics of FD communications, and presents main
application scenarios. The major resource allocation problems in
these applications are discussed in Section~III. Then we provide two example scenarios in Section~IV on
link selection for FD-MIMO networks, and user pairing and subcarrier
assignment for FD-OFDMA. In Section~V,  we
draw the main conclusions, and also discuss future research
directions.

\section{Full-Duplex Communication Basics and Applications}

In this section, we first briefly introduce FD communication systems, and then, present possible application scenarios.

\subsection{Basics of Full-Duplex Communication}

\begin{figure}[h]
\centering
\includegraphics[width=4in]{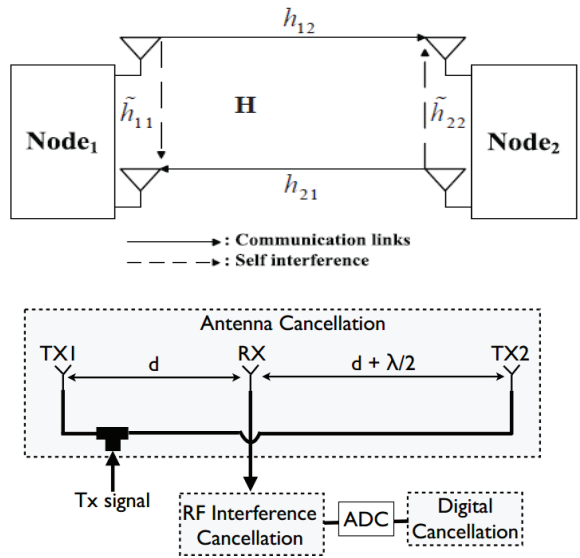}
\caption{Full-Duplex Wireless Communications and Example of Self Interference Cancelation at each node} \label{fd}
\end{figure}

Fig.~\ref{fd} presents a two-node FD
communication system with a transmit and a receive antenna at each
node. Two nodes can concurrently transmit and
receive the signals at the same frequency and time interval, which
leads to severe self interference caused by the signal leakage from
the transmit RF unit to receive RF unit, as shown in Fig.~\ref{fd}.
One way to cancel the self interference is by antenna cancellation,
and there are many analog and digital signal process techniques \cite{Melissa, fmradios, Riihonen2011TWC} developed
recently for self interference cancellation.

Depending on the distance between the transmit and receive antennas,
the self interference can be $50$-$110$dB larger than the received
signal. As a result, the signal received at each node is dominated
by the self interference. This will overwhelm the AD/DA conversion
process due to its limited dynamic range. Consequently, the
effective bits for the desired received signal is much smaller, and
the resulting SINR is low. Therefore, the self interference needs to
be mitigated before the ADC in analog circuit. After the analog self
interference cancelation, the remaining interference can be further
reduced by the active digital cancelation. However, due to the
practical constraint, the interference cannot be completely
suppressed. In the literature, depending on the self-interference
cancelation techniques, the residual self-interference~(RSI) can be
modeled as AWGN, Rayleigh or Rician distributed
variables~\cite{Emadi,XiZhang}.

As a result, the signals received at each node are a combination of
the signal transmitted by the other source, the RSI, and the noise. As shown in Fig.~\ref{fd}, at node $1$, it has
\begin{equation}\label{eq:SM_recv_fd}
{y_1} = \sqrt{p_2}{h_{21}}{x_2} +   \sqrt{p_1}{\tilde h_{11}}{x_1} + {n_1},
\end{equation}
where $p_1$ and $p_2$ represent the transmit power at each node , $h_{21}$ denotes the
communication channel from node $2$ to node $1$, $\tilde h_{11}$ represents the interference channel, and $n_1$ denotes the noise term. Thus, the instantaneous received
SINR can be calculated as
\begin{equation}\label{eq:SM_SINR_fd}
{\gamma _{1}} = \frac{{{{\left| {{h_{21}}}
\right|}^2}{p_2}}}{{{{\left| {{{\tilde h}_{11}}} \right|}^2}{p_1} +
N_0}}.
\end{equation}
From (\ref{eq:SM_SINR_fd}), it is obvious that the instantaneous
SINR decreases as the RSI increases. As indicated in (\ref{eq:SM_SINR_fd}), the values of
$|h_{21}|^2$, $|\tilde h_{11}|^2$, $P_1$, and $P_2$  have a strong impact on
SINR, and in turn will affect the system performance
significantly, while in traditional HD communication system, the major effects come from the transmit side only. Therefore, the effective resource allocation that can
further reduce the effects of RSI is crucial
for FD communication and networks. In the next sections, we will
present various resource allocation algorithms for the FD wireless
systems.

\subsection{Key Full Duplex Communications Networks}

\subsubsection{Full-Duplex MIMO Networks}

\begin{figure}[h]
\centering
\includegraphics[width=4.2in]{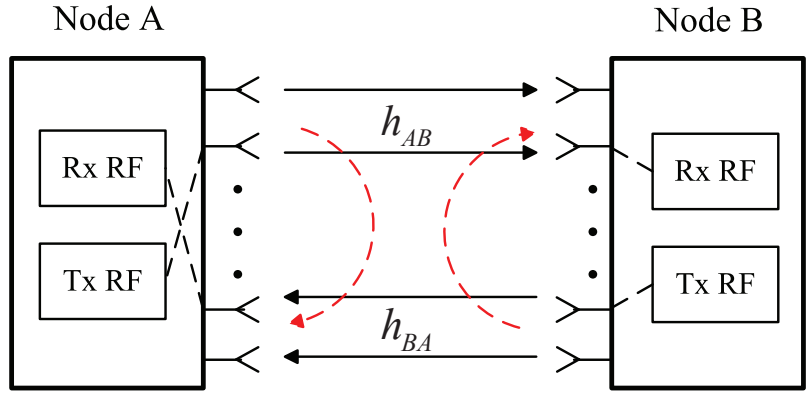}
\caption{Full-Duplex MIMO Networks}
\label{mimo}
\end{figure}

Let's first introduce a simple bidirectional communication MIMO
system by extending the setup in Fig.~\ref{fd}. Fig.~\ref{mimo} consists of a pair of FD
MIMO transceivers, where each node is equipped
with multiple antennas, respectively. Specifically, at each node, some
antennas will be used for transmission and some can be chosen for
reception. Both nodes operate in the same frequency band at the same
time~\cite{zhou2014, zhou2015}.

\subsubsection{Full-Duplex Relay Networks}

Relaying technology has been widely used in many communications
systems. Traditional relay systems, where a source node communicates
with a destination node through one or multiple relays, operates in
the HD mode. This leads to the loss of spectral efficiency because
the relay node needs to receive and retransmit the signals in the
orthogonal resources. By equipping the relay nodes with FD radios,
the relay can receive and retransmit signals simultaneously, and
thus the spectral efficiency can be improved\cite{cui2014,Suraweera2014,yang2014}. In this subsection, we
introduce FD cooperative and two-way relay networks.


\begin{figure}[h!]
\centering
\includegraphics[width=6in]{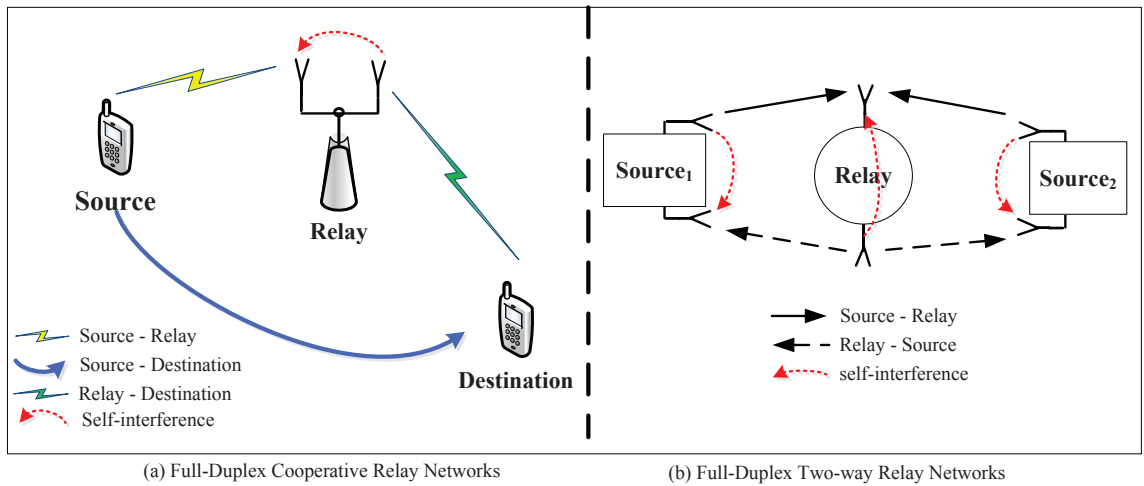}
\caption{Full-Duplex Cooperative Networks}\label{relay}
\end{figure}

\textbf{Full-Duplex Cooperative Relay Networks}: Fig.~\ref{relay}~(a) illustrates a simple FD relay network consisting
of one HD source, one HD destination, and one FD cooperative relay
node. Both the source and relay nodes use the same time-frequency
resource and the relay nodes work in the FD mode with two antennas
(one for transmission and one for reception). The communication
process can be briefly described below:
\begin{itemize}
  \item The source transmits signals to both the FD relay and destination;
  \item At the same time the FD relay forwards the signals received in the previous time slots to the destination.
\end{itemize}

As a special case, when there is no direct channel link between the
source and destination, the scenario in Fig.~\ref{relay} ~(a) can be reduced to FD
one-way relay networks.

\textbf{Full-Duplex Two-way Relay Networks} Similar to the FD one way relay network, employing the FD relaying
in two-way relay networks can also greatly improve its
performance~\cite{cui2014}. As shown in Fig.~\ref{relay}~(b), the FD
two-way relay system consists of two sources, and one relay node and
all nodes work in the FD mode with two antennas, one for
transmission and one for reception. The direct link between two
source nodes does not exist. The
communication process consists of two phases:
\begin{itemize}
  \item Two source nodes transmit signals to the relay node, while receiving the signal sent from the relay node at the same time;
  \item The relay broadcasts the signals received in the previous time slot to both source nodes and meanwhile receiving the signals from the sources.
\end{itemize}

\subsubsection{Full-Duplex OFDMA Cellular Networks}

\begin{figure}[h]
\begin{center}
\centerline{\includegraphics[width=4in]{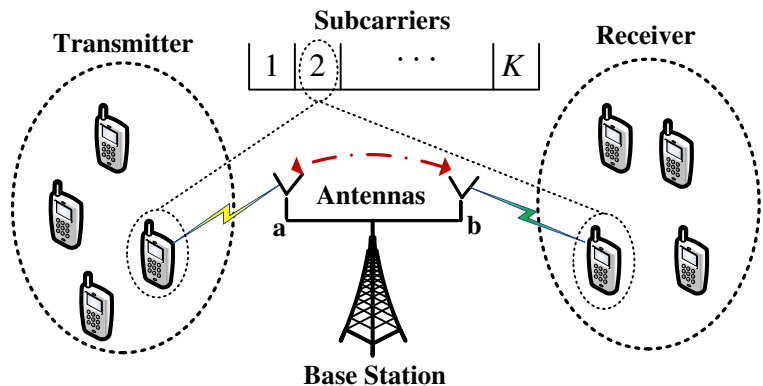}}
\caption{Full Duplex MIMO OFDMA networks} \label{ofdma}
\end{center}
\end{figure}

OFDMA has been widely used in many wireless and cellular
communication systems. Design of efficient OFDMA FD networks has
recently stimulated new research interest. As shown in
Fig.~\ref{fd}, in FD-OFDMA cellular networks, the transmit~(TX) and
receive~(RX) users need to be properly paired into separate
transceiver pairs, and each pair of TX and RX users simultaneously
communicate with the FD BS over the same subset of subcarriers.
Within each pair, the transmission of TX user will cause the
co-channel interference to the RX user, and this interference varies
largely with the mutual distance between the TX and the RX user of
each pair. The TX and RX users pairing, subcarrier and power
allocation among different pair of users need to be properly managed
to achieve the optimal sum rate performance in the network. Due to
the combinatorial nature of pairing multiple TXs, RXs, and
subcarriers, and also the complexity of optimal power allocation to
each subcarrier-transceiver pair, resource allocation in such a FD
OFDMA network can be very challenging~\cite{di2014}.

\subsubsection{Full-Duplex Heterogeneous Networks}

\begin{figure}
    \begin{center}
    \includegraphics[width=4.5in]{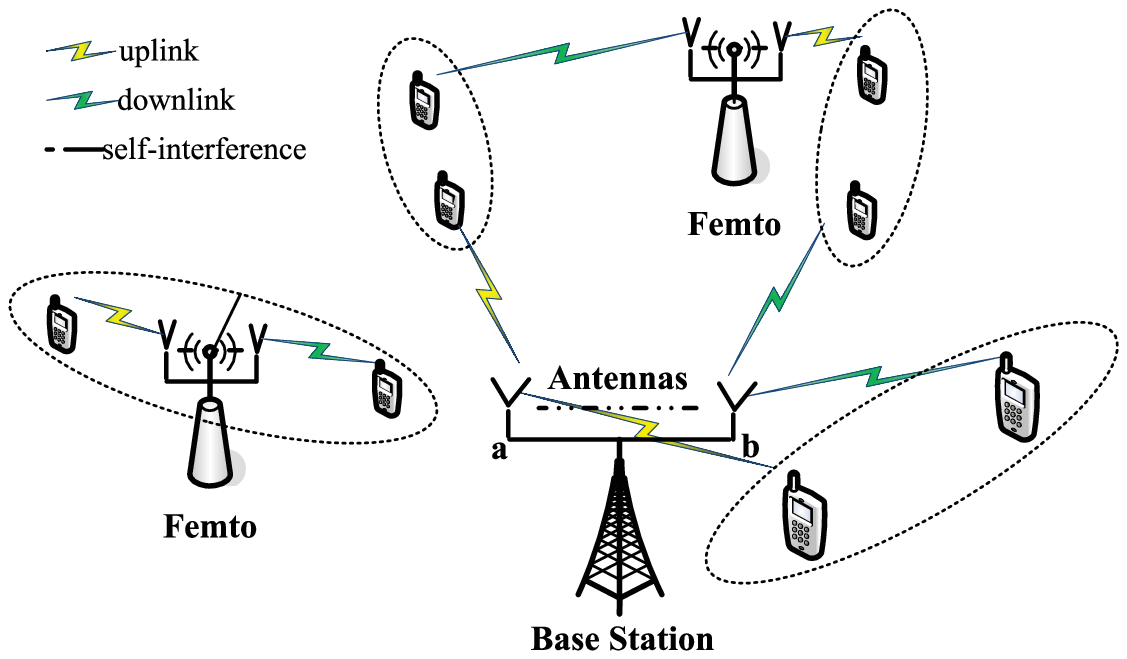}
    \caption{Full-Duplex Heterogenous Networks}
    \label{HetNet}
    \end{center}
\end{figure}

The increased number of wireless communication users have stimulated
a continuous demand for new resource allocation schemes that are
able to decrease the traffic congestion. Accordingly, two-tier
heterogeneous networks have emerged as an effective solution by
offloading the traffic from the macro BSs to the FAPs. In this subsection, we introduce the FD two-tier
heterogeneous networks, consisting of micro and small cells all of
which are employed with FD radios, as shown in Fig.~\ref{HetNet}.

A FD-HetNet consists of a single BS and multiple FAPs, all equipped
with multiple antennas. Each cell has multiple users that attempt to
connect either to the BS or the FAPs. Both BS and FAPs work in the
FD mode. Obviously, employing FD radio will increase the
interference level at each BS and FAPs because of the self interference and inter-cell interference caused by the asynchronous downlink and uplink transmissions. The users can select to connect either to
the BS or to the FAP, and there exist following four different
possible modes and different modes that need to be selected for
different users to optimize the network throughput~\cite{Sultan2014}:
\begin{itemize}
  \item Both uplink and downlink users are connected to the BS;
  \item The uplink user is connected to the BS but the downlink user is connected to the FAP;
  \item Both the uplink and downlink users are connected to the FAP;
  \item The uplink user is connected to the FAP and the downlink user is connected to the BS.
\end{itemize}

\section{Resource Allocation Problems in Full-Duplex Communications Systems}

In this section, we summarize the main resource allocation problems
for FD communication systems. FD communication provides
another dimension of resource and requires the new design of
resource allocation algorithms, but the performance is also greatly
affected by the RSI, as shown in (\ref{eq:SM_SINR_fd}). Some key
resource allocation problems for FD communications systems can be
summarized below.

\subsection{Mode Switch}

FD radio allows a node to send and receive signals at the same time
in the same frequency band. However, it is practically impossible to
have perfect self interference cancelation (the cancelation
capability also depends largely on the node's signal processing
capability), and thus the amount of RSI greatly affects the
performance of the FD system. As a result, in some scenarios, the HD
mode may outperform the FD mode for certain RSI values. On the other
hand, due the limited size of transmitter and receivers, many
wireless communication systems suffer from the spatial correlation
which degrades the performance gain of the HD mode. Taking into
account the practical RSI and spatial correlation, either FD or HD
might be optimal. To this end, it is essential to understand the
performance of these two transmission modes in order to identify the
conditions for which FD or HD performs best. This motivates the
adaptive mode switching between the FD and HD modes based on the RSI
and channel conditions to maximize the ergodic capacity~\cite{Yao2015}.

\subsection{Power Control}

Power control has been a commonly used approach in multi-user
communication systems to optimize system performance such as link
data rate, network capacity and coverage. Unlike traditional
wireless networks, FD communication suffers from the RSI, and thus,
increasing transmit power can improve the signal strength in the
receiver side, but on the other side increases the RSI at its own
receiver, as indicated in (\ref{eq:SM_SINR_fd}). Therefore, due to
the existence of RSI, corresponding power control algorithm needs to
be properly redesigned in order to maximize system performance of
all users. The different power constraints, e.g. {\em total} or {\em
individual} transmit power, will lead to different designs and final
solutions~\cite{yang2014}. Moreover, as detailed below, different FD systems require
different power control algorithms:

\begin{itemize}
  \item FD-MIMO: In bidirectional FD-MIMO communication, the antennas at the FD node are divided into transmit and receive antenna sets, and water-filling power allocation can be applied at the transmit antenna set to maximize the sum rate based on individual power constraint. The water-filling power allocation at each nodes needs to take into account the self-interference from the transmit antenna set to the receive antenna set at each node and the power pouring results at each node can be significantly different;
  \item FD-Relay: In FD-Relay networks with individual power constraint at each relay, the relayed signals are corrupted by the RSI, and thus, the received signals at the destination are the combination of the desired signals plus RSI introduced at the relay node. Increasing the transmit power at the relay will increase the power of desired signal at the destination, but on the other side it will increase the RSI in the destination received signal. Therefore, there is an optimal transmit power at the relay to maximize the performance at the receiver node;
  \item FD-OFDMA: For FD-OFDMA with one FD BS and HD multiple users, uplink (transmit) and downlink (receive) users are paired to communicate with the FD BS at the same time. The transmit power can be allocated at the BS side with total power constraint by splitting the power among all the subcarriers for different user pairs. At the user side, power control needs to take into account the inter-user distance among the transmit-receiver user pair. Thus, the optimal power control in the FD-OFDMA system depends on many factors, which requires multi-dimensional optimization for the BS and the mobile users\cite{di2014};
   \item FD-HetNet: Similar to FD-OFDMA, the power control can be performed for the FD BS and FAPs and HD users in FD-HetNet to optimize the network performance. However, both the inter-cell interference and RSI need to be considered jointly in optimizing the overall network performance.
\end{itemize}

\subsection{Transmit Beamforming}

Transmit beamforming is a general signal processing technique used
to control the directionality of transmission in order to provide
a large antenna array gain in the desired directions, and has been
widely applied in the 3rd and 4th generation wireless
communication systems. To change the directionality of the array
when transmitting, a beamformer controls the phase and relative
amplitude of the signal at the transmitter in order to create a pattern
of constructive and destructive interference in the wave front. For FD
communications, it would be greatly beneficial to design the robust
transmit beamforming algorithms that can improve the signal strength
at the receiver side, and meanwhile reduce the self interference
subject to various design criteria such as the minimum mean square
error. Below are the discussions of beamformer design for
different FD systems.

\begin{itemize}
  \item FD-MIMO: In this case, the transmit antenna set at each FD nodes can perform transmit beamforming to simultaneously transmit information and reduce the interference to its own received signals. The design is to jointly optimize the system sum rate.
  \item FD-OFDMA: In FD-OFDMA, the FD BS is equipped with multiple antennas, consisting of transmit and receive antenna sets, while the users only operate in the HD transmission mode due to hardware constraint. In this case, the BS can construct beamformer to support multiple users in the downlink while maximizing the received SINRs at BS by minimizing the RSI. Besides, the beamforming design also needs to consider appropriate pairing of downlink and uplink users, which also significantly affect the self interference at the BS, and the co-channel interference among downlink and uplink users.
\end{itemize}

\subsection{Link Selection}

For a FD communication system, each antenna can be configured to
transmit or receive the signals. This will create multiple possible
virtual links between two nodes, with one virtual link representing
the channel from a transmit antenna of one node to a receive antenna
of the other. Since the FD radio enables simultaneous bidirectional
information exchange between two nodes, an important question arisen
in such a scenario is how to optimally select the link for each
direction to optimize the system performance. Obviously, the optimal
selection requires exhaustive search among from all possible antenna
links. However, as the number of antennas increases, such a
brute-force search suffers from very high computational complexity,
and thus, simple but near-optimal selection algorithms need to be
developed. Next let us discuss some selection algorithms for
the FD-MIMO, FD-Relay, and FD-HetNet systems.

\begin{itemize}
  \item Antenna selection in FD-MIMO systems: For FD-MIMO systems with $N_A$ and $N_B$ antennas equipped at two nodes. Such a FD MIMO system will create $N_A\times N_B$ possible virtual links between two nodes, with one virtual link representing the channel from a transmit antenna of a node to a receive antenna of the other node. The challenge is to design the simple but (near) optimal bidirectional link selection algorithm to optimize the bidirectional sum rate or sum symbol error rate (SER)~\cite{zhou2014, zhou2015}.
  \item Joint antenna and relay selection in FD-Relay systems: In a general FD relay networks consisting of one source, one destination, and $N$ FD relays, and each antenna of the FD relay is able to transmit/receive the signal. Hence, the source, relays, and destination form a number of  virtual end-to-end links. In this case, each antenna of the FD relay is able to transmit/receive the signal. Each relay adaptively selects its TX antenna and RX antenna based on the instantaneous channel conditions, and the optimal single relay with the optimal TX/RX antenna configuration is selected to maximize the end-to-end SINR of the system~\cite{cui2014,yang2014}.
  \item Coordinated multiple point transmission: In FD HetNet, the users can select to communicate with the transmit/receive antenna of multiple available FD access points (BSs or FAPs) in a coordinated way so as to form the joint multiple point transmission, improving  spatial and frequency utilization efficiency~\cite{Sultan2014}.
\end{itemize}

\subsection{Subcarrier Allocation}

In an OFDMA system, at each time slot, disjoint sets of subcarriers can
be assigned to different users based on some target objectives. The
users then in turn transmit data by spreading the information across
the assigned subcarriers. In a FD-OFDMA network consisting of one FD
BS with $N_f$ subcarriers, $N_u$ uplink users, and $N_d$ downlink
users, a fundamental challenge arisen in such system is how to pair
uplink and downlink users, and allocate subcarrier across these user
pairs, in order to optimize the network performance. The subcarrier
allocation involves allocating the different subsets of subcarriers
to different users by taking into account the RSI at the BS and the
co-channel interference between the uplink and downlink users within
each user pair~\cite{di2014}. This is significantly different from the traditional
subcarrier allocation problem and present further research
challenges in resource allocation.

Similarly, in FD-Relay networks, consisting of multiple source and
destination nodes, and FD relay nodes using OFDM transmission, the
corresponding subcarriers should be properly allocated at the relay
for different source-destination pairs.

\section{Resource Allocation Examples}
In the following two subsections, we provide two example scenarios
to illustrate how to conduct FD resource allocation.
\subsection{Bidirectional Antenna Link Selection for Full-Duplex MIMO Networks}

In this section, we discuss in details the specific link selection
problem and its near-optimal algorithms for FD-MIMO systems~\cite{zhou2014,zhou2015}. We
consider a bidirectional communication scenario between a pair of FD
transceivers, nodes $A$ and $B$, as illustrated in Fig.~\ref{mimo},
where nodes $A$ and $B$ are equipped with $N_A$ and $N_B$ antennas,
respectively. Both nodes use the same frequency band at the same
time for FD operation. Each node employs only one transmit and one
receive RF chain, and any antenna can be configured to connect
either transmit or receive RF chains. In the proposed simultaneous
bidirectional link selection scheme, two antenna links are selected
for simultaneous bidirectional communications by choosing a pair of
transmit and receive antenna at both ends for each direction. Within
each antenna pair, one antenna is selected for transmission and one
is for the reception to maximize the sum rate~(Max-SR) or minimize
the sum symbol error rate (Min-SER), respectively.

\subsubsection{Maximum Sum-Rate (Max-SR)}

In the Max-SR selection criterion, two communication links from the $I_T$-th transmit antenna at node A
to the $J_R$-th receive antenna at node B and the $J_T$-th transmit
antenna at node B to the $I_R$-th receive antenna at node A, are
selected to maximize the bidirectional sum rate~\cite{zhou2014}. This is equivalent to select, among all the possible
antenna configurations, the optimal transmit-receive antenna pairs,
denoted by $(I_T, I_R)$ at node $A$ and $(J_T, J_R)$ at node $B$, to
maximize the sum rate
\begin{equation}\label{eq:AS_maxsr}
\left\{ {({I_T},{J_R}),({I_R},{J_T})} \right\} =
\mathop {\arg \max }\left\{ {{\mathrm{Rate}}\left( { {\gamma^{{i_t}{j_r}}}} \right)} { + {\mathrm{Rate}}\left( {{ {
{{\gamma^{{i_r}{j_t}}}} }} }\right)} \right\}.
\end{equation}
We use $\mathrm{Rate}(\cdot)$ represents transmission rate, ${\gamma^{{i_t}{j_r}}}$ and ${\gamma^{{i_r}{j_t}}}$ to denote the corresponding instantaneous SINR of the selected transmit-receive antenna pairs of nodes $A$ to $B$ and nodes $B$ to $A$, respectively.

\subsubsection{Minimum Sum-SER (Min-SER)}

In the Min-SER selection criterion, the bidirectional antenna links are selected to minimize the sum SER,
\begin{equation}\label{eq:AS_minser}
\left\{ {({I_T},{J_R}),({I_R},{J_T})} \right\} = \mathop {\arg \min
}
\left\{ {\mathrm{SER}\left( {\gamma^{{i_t}{j_r}}}\right)} { +
\mathrm{SER}\left( {\gamma^{{i_r}{j_t}}}\right)} \right\},
\end{equation}
where $\mathrm{SER}(\cdot)$represent the SER.

\begin{figure}[h]
\centering
\includegraphics[width=5.5in]{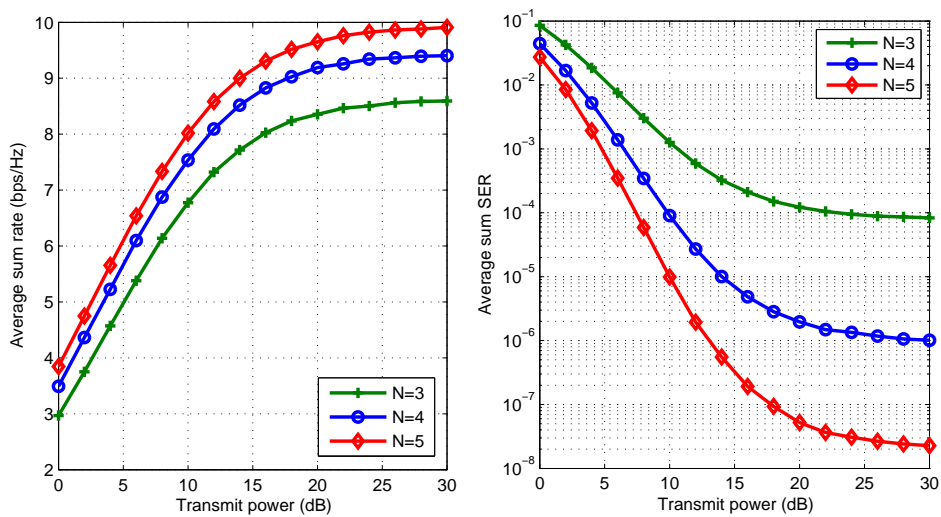}
\caption{System performance of the Max-SR and Min-SER methods, where $N=3,4,5$.}
\label{simulation_fd_mimo_as}
\end{figure}

Fig.~\ref{simulation_fd_mimo_as} illustrates the average sum rate of the Max-SR and Min-SER for different numbers of antennas $N=3,4,5$. It can be observed that the average sum rate increases with the number of antennas, while the the average sum SER performance decreases with the number of antennas.

\subsection{User and Subcarrier Pairing for Full-duplex OFDMA Networks}

In this subsection, we consider a FD-OFDMA cellular network, consisting a FD BS, and multiple TX and RX users and discusses the resource allocation issues. The contemporary matching theory is proposed to solve the complicated resource pairing problems~\cite{di2014}.
As shown in Fig.~\ref{ofdma}, in FD-OFDMA cellular networks, the BS
needs to allocate a subset of non-overlapping subcarriers to each
pair of users, so that each pair of users and the BS form a FD
transceiver unit, in which one user acts as a TX and the other acts
as a RX. Note that each subcarrier is assigned to a user pair only.
Without loss of generality, we assume that the BS is equipped with
two antennas $a$ and $b$, and multiple users each with one antenna.

To facilitate describing the user and subcarrier pairing process, define a three-dimensional $M \times M \times K$ pairing matrix $\textbf{X} = \{ 0,1\}$, where ${x_{k,i,j}} = 1$ denotes that $T{X_i}$ and $R{X_j}$ are paired and use subcarrier $k$. Our objective is to maximize the sum-rate of the system by jointly optimizing the pairing variables $\{ {x_{k,i,j}}\} $ and the power variables $\{ p_{s,j,k}\}$. The optimization problem can be formulated as:
\begin{align} \label{joint_optimization}
&\mathop {\max } \sum\limits_{k = 1}^K {\sum\limits_{i = 1}^M {\sum\limits_{j = 1}^M {\mathrm{Rate}(\{x_{k,i,j}\},} } } \{p_{s,j,k}\})\nonumber\\
\emph{s.t.}&\quad \mbox{each TX can only be paired with one RX, and vice versa,}\nonumber\\
&\quad \mbox{each subcarrier can only be assigned to one transceiver unit, and vice versa,}\nonumber\\
&\quad \mbox{the total transmit power of the BS is subject to its peak power constraint $P_s$}.
\end{align}

The matching algorithm can be briefly described as follows. First, define a price for each SR unit and set the price to zero. These prices are fictitious money without any physical meanings that are considered as the matching cost for each TX. The price of any SR unit, $SR_{k,j}$ is the sum of $R{X_j}$'s price and subcarrier $k$'s price. Then in each step, any $T{X_i}$ that is still not matched proposes to its most preferred $SR_{k,j}$ according to the achieved sum-rate and the cost of the corresponding $T{X_i}$-$SR$ unit. If $R{X_j}$ or subcarrier $k$ receives offers from more than two TXs, they increase their prices with a price step number until only one offer is received. When $R{X_j}$ and subcarrier $k$
both receive only one offer, which comes from $T{X_i}$, they will be matched together. The matching algorithm is iterative and ends if all the TXs are matched and no new offer is being made. This point is called the \emph{equilibrium point} of the matching, which also indicates that the convergence has been achieved.

\begin{figure}[h]
\begin{center}
\centerline{\includegraphics[width=5.5in]{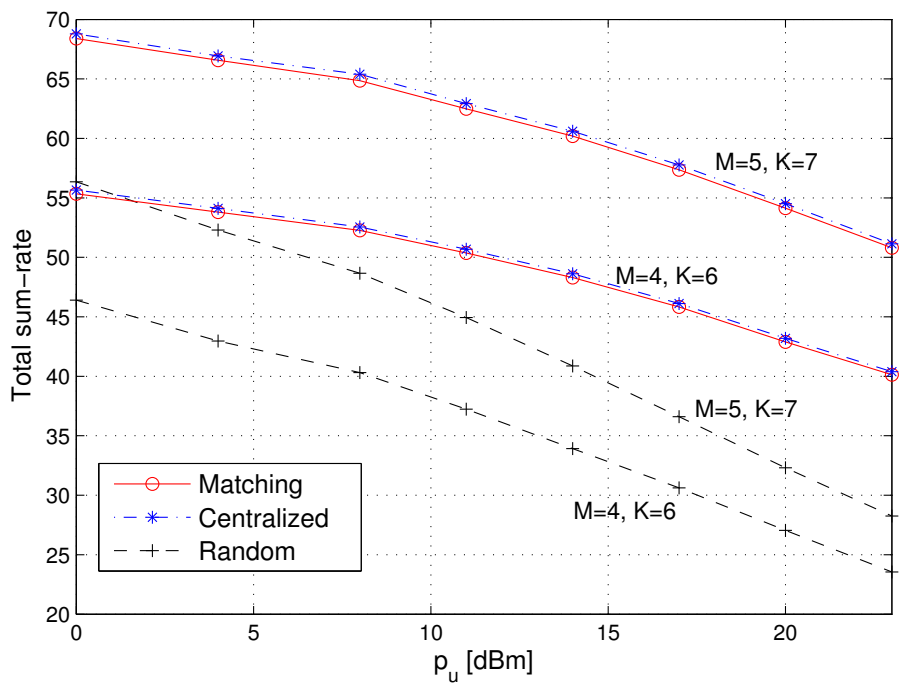}}
\caption{Sum rate performance in terms of transmitted power of each user.} \label{simulation_fd_ofdma}
\end{center}\vspace{-8mm}
\end{figure}
To evaluate the performance, the centralized solution is compared. Besides, a random matching algorithm is utilized, in which the TXs, the RXs and the subcarriers are randomly matched with each other while satisfying the system constraints. These two algorithms are considered as the upper and lower bound solutions in terms of the complexity. Fig.~\ref{simulation_fd_ofdma} illustrates the total sum-rate vs. transmitted power of each user. It shows that the proposed matching algorithm provides a total sum-rate of the network quite close to the centralized algorithm. Besides, the complexity level of the proposed algorithm is much lower than that of the centralized algorithm such that when $M = 5$, and $K = 7$, the number of iterations in the centralized algorithm is 7200, which is 928.57$\%$ higher than that in the proposed algorithm which is no more than 700. Also, the proposed matching algorithm performs significantly better than the random matching.

\section{Conclusions and Future Research Directions}

This article presented the recent development of FD signal
processing and network protocols by considering representative FD
communications networks: FD-MIMO, FD-Relay, FD-OFDMA, and FD-HetNet
networks. The associated resource allocation problems in these
systems are discussed, including the mode switch, power control,
link selection and pairing, interference-aware beamforming and
subcarrier assignment. Then we illustrate two example FD resource
allocation scenarios. Since the FD communication creates multiple
possible virtual links between each node, we present simultaneous
link selection for FD-MIMO networks by Max-SR and Min-SER criteria;
Besides, We also elaborate how matching theory can be applied to
solve the user and subcarrier pairing problems in FD-OFDMA cellular
networks.

FD communication is a very promising technology, and there are many potential future research directions in this area, for example,
\begin{itemize}
  \item FD cognitive radio networks: In traditional cognitive radio networks, secondary users (SUs) typically access the spectrum of primary users by a two-stage ``listen-before-talk'' protocol, i.e., SUs sense the spectrum holes in the first stage before transmit in the second. With a FD radio, it allows SUs to simultaneously sense and access the vacant spectrum. As a result, research topics such as spectrum sensing algorithms, dynamic spectrum access, communication protocol design, etc, need to be redeveloped~\cite{Liao2014};
  \item Physical-layer security: Physical layer security provides an alternative security solution by considering the physical characteristics of wireless links. Traditional opinion typically treats the interference as a disadvantage, but in physical-layer security, interference can be utilized to interferer the malicious nodes. In FD communication, the self-interference can be certainly reused to improve the network secrecy capacity. Various research topics such as secrecy capacity analysis, power control, beamforming, etc., are worth of further investigation.
\end{itemize}

\end{document}